\documentclass[prb,preprint,preprintnumbers,amsmath,amssymb]{revtex4}

\usepackage{graphicx}

\usepackage{dcolumn}

\usepackage{bm}

\begin{document}


\title{Drift and Diffusion of Spins Generated by the Spin Hall Effect}

\author{N. P. Stern}
\author{D. W. Steuerman}
\author{S. Mack}
\author{A. C. Gossard}
\author{D. D. Awschalom}
\email{awsch@physics.ucsb.edu}

\affiliation{Center for Spintronics and Quantum Computation,
University of California, Santa Barbara, CA 93106 USA}

\date{\today}

\begin{abstract}

Electrically generated spin accumulation due to the spin Hall effect
is imaged in $n$-GaAs channels using Kerr rotation microscopy,
focusing on its spatial distribution and time-averaged behavior in a
magnetic field. Spatially-resolved imaging reveals that spin
accumulation observed in transverse arms develops due to
longitudinal drift of spin polarization produced at the sample
boundaries. One- and two-dimensional drift-diffusion modeling is
used to explain these features, providing a more complete
understanding of observations of spin accumulation and the spin Hall
effect.

\end{abstract}

\maketitle

The observation of current induced spin
polarization\cite{Kato:2004a} and the spin Hall
effect\cite{Kato:2004b} demonstrate all-electrical generation of
spin polarization through the spin-orbit interaction. A detailed
understanding of spin accumulation produced by these phenomena is a
prerequisite to advancing spin-based electronics applications.
Sample boundaries in spin-orbit systems are an active topic of
research, affecting both the observed spin accumulation near and
away from sample edges \cite{Sih:2006} and even the proper
theoretical definition of a spin current\cite{Shi:2006}. For spin
Hall systems in the extrinsic regime, many of these complications
can potentially be avoided because the spin-orbit coupling is weak,
enabling simple solutions to spin diffusion equations\cite{Tse:2005}
which have been used to explain spin accumulation in a number of
studies \cite{Kato:2004b, Valenzuela:2006, Sih:2006, Stern:2006}.

Despite these expected simplifications in the analysis of the
extrinsic spin Hall effect, spatially-resolved optical measurements
have revealed spin accumulation which is not entirely explained by
simple one-dimensional spin diffusion models\cite{Kato:2004b,
Sih:2006}. In this Letter we measure spin accumulation with
unambiguous contributions from both drift along electric fields and
spin precession about an applied magnetic field. We develop a
diffusion model in one- and two-dimensions which can reproduce the
observations as well as allow reinterpretation of the details of
recent experiments on the spin Hall effect.  These results provide a
more complete understanding of the behavior of electrically
generated spin polarization in micron-scale devices.

The structures discussed in this manuscript are fabricated from a
2-$\mu$m thick epilayer of $n$-GaAs deposited on 200 nm of undoped
Al$_{0.4}$Ga$_{0.6}$As grown by molecular beam epitaxy. The $n$-GaAs
is Si-doped with $n = 3\times 10^{16}$ cm$^{-3}$ (at 30 K) for long
spin coherence times as in Ref. \onlinecite{Kato:2004b}. Using
standard photolithography and wet etching, the $n$-GaAs is patterned
into 300-$\mu$m long, 60-$\mu$m wide channels with a 60-$\mu$m long,
30-$\mu$m wide transverse arm (Fig. 1a).  The $n$-GaAs is contacted
by annealed Au/Ni/Au/Ge/Ni. The devices are mounted in a helium flow
cryostat with the channel defined to be the $x$ direction and a
magnetic field $B$ applied in the $y$ direction.  All measurements
reported are at 30 K, and have been repeated in a variety of similar
$n$-GaAs wafers and channels of various dimensions. An AC square
voltage with frequency $f_V = 1.168$ kHz is applied across the
device, driving a current in the $\pm x$ direction and generating
spin polarization $s^z$ in the $\pm z$ direction due to the spin
Hall effect\cite{Kato:2004b}.

Spin accumulation is measured using a Kerr rotation microscope with
a spatial resolution of $\sim 1$ $\mu$m\cite{Stephens:2003}.  A
linearly polarized beam from a pulsed Ti:sapphire laser tuned to
$\lambda = 821$ nm is focused through a microscope objective normal
to the sample surface and the polarization of the reflected beam is
analyzed with a balanced photodiode bridge and lock-in amplifier
operating at $f_V$.  The axis of linear polarization rotates by an
angle $\theta_{K}$ proportional to $s^z$ due to the Kerr effect. The
microscope objective is scanned over the device to spatially map
$s^z$ while using the reflectivity of the laser over the channel
edges to calibrate the coordinate position.

Spatial profiles across the channel width (Fig. 1b) show a
hyperbolic sine form for $s^z(y)$, as expected from solution of
basic diffusion equations\cite{Kato:2004b, Tse:2005}. Spin
accumulation in the channel is measured at least a few spin
diffusion lengths ($L_s \sim 5 \mu$m) away from the arm edge to
minimize diffusive effects from the transverse arm. Sweeps of $B$
produce a quasi-Lorentzian lineshape due to the time-averaged
projection of $s^z$ known as the Hanle effect\cite{Kato:2004b}(Fig.
1c). The spin coherence time can be extracted from the half-width
$B_{1/2}$ of a Lorentzian fit to the magnetic field scans, $\tau =
\hbar / g \mu_B B_{1/2} $, where $\hbar$ is the reduced Planck's
constant, $g = -0.44$ is the g factor for GaAs, and $\mu_B$ is the
Bohr magneton. $\tau$ decreases with increasing applied $E$, which
has been previously explained as increased D'yakonov-Perel spin
scattering due to electron heating\cite{Beck:2006}.  As a function
of $y$, the central peak narrows away from the channel edge, causing
an apparent increase in the extracted $\tau$ toward the center of
the channel (Fig. 1b inset). This feature is consistent with
previous observations of the spin Hall effect\cite{Kato:2004b,
Stern:2006}. Increased averaging in the current measurements also
reveals that $s^z$ changes sign away from the central peak.
Similarly, magnetic field scans in the transverse arm are more
complicated than those in the main channel, exhibiting sign
reversals and narrow central peaks (Fig. 1d) that cannot be fit by a
Lorentzian lineshape. The cause of the departures from simple
Lorentzian behavior in both the channel and the arms is expected to
be due to a combination of spin precession, drift, and diffusion
\cite{Sih:2006}, but detailed calculations of both the spatial and
$B$ dependences of the experimental data have so far been lacking.
The remainder of this Letter will present modeling necessary to
understand the above observations of spin accumulation due to the
spin Hall effect.

In order to model spin accumulation, we assume that spin is
generally conserved except for a decay with time constant $\tau$.
The sample boundaries are treated as hard walls; neither charge nor
spin can penetrate sample edges, and the edge causes no spin-orbit
induced spin-flips. These approximations are both reasonable as GaAs
is expected to be in the extrinsic limit of the SHE where spin-orbit
coupling is weak. Under these assumptions, the vector spin density
$s^i$ obeys a steady state continuity equation with additional spin
precession and decay terms:
\begin{equation}
\frac{d s^i}{dt} = -\nabla \cdot \vec{J^{i}_{s}} - \frac{s^i}{\tau}
+ (\vec{\omega} \times \vec{s})^i = 0\label{eq:1}
\end{equation}
where $\vec{\omega} = g \mu_B \vec{B/ \hbar}$ is the Larmor
frequency vector and $\vec{J^{i}_{s}}$ is the spin current of the
$i$-th component of spin, which is defined in the simplest fashion
as a sum over all particles of the product of spin and velocity:
$\vec{J^{i}_{s}} = \sum s^i \vec{v}$. A general treatment using a
more complicated definition of spin current\cite{Shi:2006}, which is
potentially necessary in the intrinsic regime, is not necessary in
this case. For electrons with isotropic, spin-independent transport
properties, the spin current tensor is:
\begin{equation}
J^{i}_{s,j} = -\mu s^i E_j - D \partial_{j} s^i + \sigma_{SH}
\epsilon^{i j k} E_k \label{eq:2}
\end{equation}
where $\mu$ is the charge mobility, $\vec{E}$ the electric field,
$D$ the diffusion constant, $\sigma_{SH}$ the spin Hall
conductivity, and $\epsilon^{i j k}$ the antisymmetric symbol.  The
first term represents drift of spin polarized carriers along the
electric field, the second term governs diffusion of accumulated
spin polarization, and the third is the spin current generated
transverse to an electric field by the spin Hall effect.

Spin and charge coupling\cite{Tse:2005} is assumed to be negligible
for simplicity, decoupling the electric field solution from the spin
accumulation.  This additional coupling does not appear to be
necessary to capture the salient features of the experimental
observations. Introducing the spin diffusion length $L_s = \sqrt{D
\tau}$, the drift-diffusion equation in the channel is obtained from
Eq. \ref{eq:1} and Eq. \ref{eq:2}:
\begin{equation}
\partial^2 s^i +\frac{\mu \tau}{L_{s}^{2}} E_j \partial_j s^i - \frac{s^i}{L_s} + \frac{\tau \vec{\omega}}{L_{s}^2} \times \vec{s} = 0  \label{eq:3}
\end{equation}
In Eq. 3, we assume the system is in the linear transport regime,
where $\sigma_{SH}$ is independent of electric field and the
electron density is uniform. Because the electric field has no curl,
the divergence of the spin Hall current is identically zero:
$\partial_j (\sigma_{SH} \epsilon^{i j k} \vec{E_k}) = \sigma_{SH}
(\epsilon^{i j k} \partial_j E_k) = 0 $. For linear transport, the
spin Hall current does not act as a source for spin accumulation in
bulk; all spin accumulation arises from boundary accumulation at the
channel edge. Recent modeling\cite{Finkler} has shown that bulk spin
accumulation can indeed occur in a non-linear transport regime where
the divergence of the spin Hall current is not identically zero; we
do not study this effect in this manuscript.  Eq. \ref{eq:3}
represents a system of coupled linear differential equations for
$s^i$ written entirely in terms of experimentally determined
parameters. These equations include drift and spin precession terms,
both of which have demonstrable consequences in the experimental
data presented in Fig. 1.

In the case of an infinitely long channel oriented along the $x$
direction(which is applicable far from the contacts and transverse
arms), $s^i$ will be independent of $x$ and the local electric field
will be a constant $\vec{E} = E \hat{x}$, causing the drift term to
vanish and validating the assumption $\sigma_{SH}(E) = \sigma_{SH}$
of Eq. \ref{eq:3}.  Specializing to the case of $B$ along $y$, Eq.
\ref{eq:3} reduces to two non-trivial coupled differential equations
for $s^x(y)$ and $s^z(y)$, which can be solved exactly by enforcing
the boundary condition $\hat{n} \cdot \vec{J^{i}_{s}} = 0$. These
diffusion equations including spin precession were solved for the
case of spin accumulation in a one-dimensional system with polarized
spins injected from ferromagnetic contacts\cite{Johnson:1988}. The
difference in this case is that the spin polarization is generated
at the channel edge by the spin Hall effect. The spatial profile
$s^z(y)$ at $B = 0$ is the same hyperbolic sine solution used in
earlier analysis of the SHE \cite{Yui:2004b, Tse:2005}. Near the
channel edge, $s^z(B)$ has the expected Lorentzian shape from the
Hanle model. Away from the channel edge, spin precession modifies
the steady state solution $s^z(B)$, resulting in a narrowing central
peak and oscillations in the tails. Calculations of the $s^z(B)$
dependence are shown in Fig. 2a, assuming $\tau = 2.5$ ns and $L_s =
4.5 \mu$m, overlaid with the normalized data from Fig. 1b. The
narrowing of the central peak and the sign change of $s^z$ are
evident both in these $B$ scans and and the full $B$ and $y$
dependence of the data and calculations in Fig. 2c. Within the
diffusion model, the central peak narrowing and the non-Lorentzian
tails of Fig. 1c can be entirely understood as spin precession of
diffusing spins with a constant $\tau$ without invoking a
position-dependent spin coherence time.

We apply the model discussed above to two-dimensional geometries as
well, focusing on the transverse arm originally investigated by Sih
et al.\cite{Sih:2006} In this geometry, the drift term of the
equations plays an important role. Two-dimensional finite element
modeling is used to first solve for the electric field $\vec{E}(x,
y)$ and then use Eqs. 1 and 2 to find the magnitude of spin
polarization $s^z (x, y)$. As before, we fix $L_s = \sqrt{D \tau}$
to be equal to that measured from magnetic field scans along $y$ in
Fig. 1b.  Since the magnitude of the electric field, $E$, varies
throughout the sample, we use in the model the $\tau (E)$ and $\mu
(E)$ measured separately for this sample. We note that good
agreement with the data is also observed by fixing $\tau$ and $\mu$
to their $E = 0$ values, with the reduced mobility being countered
by the factor of five increase in $\tau$.  The results we present
here use the measured $\tau (E)$ and $\mu (E)$, however, as this
more likely describes the true dynamics of the mobile electron
spins.  The model accounts for both the $B=0$ spatial profile and
the field dependence observed in linecuts at $x=0$ (Fig. 2d).
Contrary to the model in Ref. \onlinecite{Sih:2006}, this model does
not require electron spins to have a drift velocity transverse to
the electron current.

In order to confirm the effects of drift in accumulation from the
spin Hall effect, we obtain two-dimensional maps of spin
polarization in the transverse arm by modulating the magnetic field
at $f_B = 3.3$ Hz and detecting the resulting Kerr rotation with
lockin amplifiers at $f_V \pm 2 f_B$. The resulting signal will be
proportional to the second derivative of the magnetic field profile
near $B = 0$\cite(Kato:2004b), allowing efficient two-dimensional
imaging of spin accumulation without time-consuming magnetic field
scans.  Fig. 3a shows an image of KR in the 60-$\mu$m long
transverse arm extending from a 300-$\mu$m long channel. Spin
accumulation is observed across the entire width of the arm,
extending 40 $\mu$m from the main channel, as observed in Ref.
\onlinecite{Sih:2006}. This measurement is performed with a
symmetric square wave current, so the observed accumulation is
symmetric about the center of the arm at $x$ = 0 (Fig. 3a).
Conclusive evidence of the importance of drift in the spin
accumulation is obtained by applying an AC voltage with a DC offset
to create a square wave electric field between 0 mV/$\mu$m and $E$.
This slight modification allows the lock-in detection technique to
remain unchanged, but removes the symmetric current flow so that
electrons only flow in one general direction along the channel.
Figs. 3b and 3c show the data and model for the spin accumulation in
the transverse arm under unidirectional current flow.  The
accumulation profile in the transverse arm is asymmetric about the
arm center, and reverses direction when the current direction is
reversed. The spin polarization observed in the transverse arm is
neither generated in the arm itself, nor does it come from the spin
Hall current generated in the bulk of the channel; it consists
primarily of spin polarization sourced near the channel edge that
travels longitudinally along the electric field lines that penetrate
into the arm. These results accentuate the importance of including
spin drift along the electric field in analysis of both one- and
two-dimensional spin accumulation profiles.

In conclusion, we have provided experimental evidence that, in
addition to diffusion, drift and spin precession are important
factors in understanding the spin polarization generated by the spin
Hall effect. Spatial profiles of spin accumulation and its evolution
in an external magnetic field can be predicted by drift-diffusion
modeling independently of any detailed treatment of the microscopic
physics of the spin Hall effect. The practical ability to accurately
model spin accumulation generated from the spin Hall effect enables
reliable engineering of electrically-generated spin accumulation
profiles in devices.

We thank NSF and ONR for financial support. N. P. S. acknowledges
the support of the Fannie and John Hertz Foundation, and S.M.
acknowledges the NDSEG Fellowship.  We gratefully thank J. Heron for
his assistance with transport measurements.

\newpage


Figure captions\\

Figure 1. (a) Schematic of the 60-$\mu$m wide channel with a transverse arm. The origin is defined to be at the center of the transverse arm in $x$ and at the center of the main channel along $y$ (red dot), and the dotted lines represent the location of scans discussed in the text.  (b) $\theta_K (B)$ taken away from the transverse arm for $x = \sim 50 \mu$m and $y = -26 \mu$m (black circles), $-22 \mu$m (black circles), and $-18 \mu$m (magenta squares).  (c) $s^z(y)$ for $B=0$ away from the transverse arm ($x = \sim 50 \mu$m).  The inset shows the spin coherence time $\tau(y)$ extracted by assuming a Lorentzian lineshape in the Hanle model. (d) Characteristic $\theta_K (B)$ data from the center of the transverse arm at $x = 0$. The line is a guide to the eye. \\

Figure 2. (a) Normalized $\theta_K(B)$ from Fig. 1c, corresponding to data near the channel edge, $\sim L_s$ from the edge, and $\sim 2 L_s$ from the edge.  Lines are calculations from the diffusion model described in the text. (b) $\tau (E)$ and $\mu (E)$ used in the model calculations.  $\tau$ is obtained from $s^z (B)$ scans near the channel edge for various $E$, while $\mu (E)$ comes from measurements on a Hall bar. (c)  Data is an image $\theta_K (B, y)$ along the main channel width clearly showing the decreasing central peak width away from channel edge and sign change away from the central peak.  The lower panel shows model calculations of $s^z (B, y)$ as discussed in the text. (d) Data is an image $\theta_K (B, y)$ at $x = 0$ in the transverse arm showing the complicated spin accumulation profile.  The lower panel shows diffusion model calculations of $s^z (B, y)$ as discussed in the text.\\

Figure3. Images of $\theta_K$ and calculated $s^z(B = 0)$ in the
transverse arm. The top panels are data and the bottom panels are
model calculations as described in the text. The electric field in
the channel has magnitude $E = 11.1 mV/ \mu$m along $\hat{x}$, and
is applied as (a) a square wave, (b) an offset square wave between 0
and $E$, (c) an offset square wave between 0 and $-E$.  Data is
normalized to the peak values for part (a).

\begin{figure}
  \begin{center}
    \includegraphics{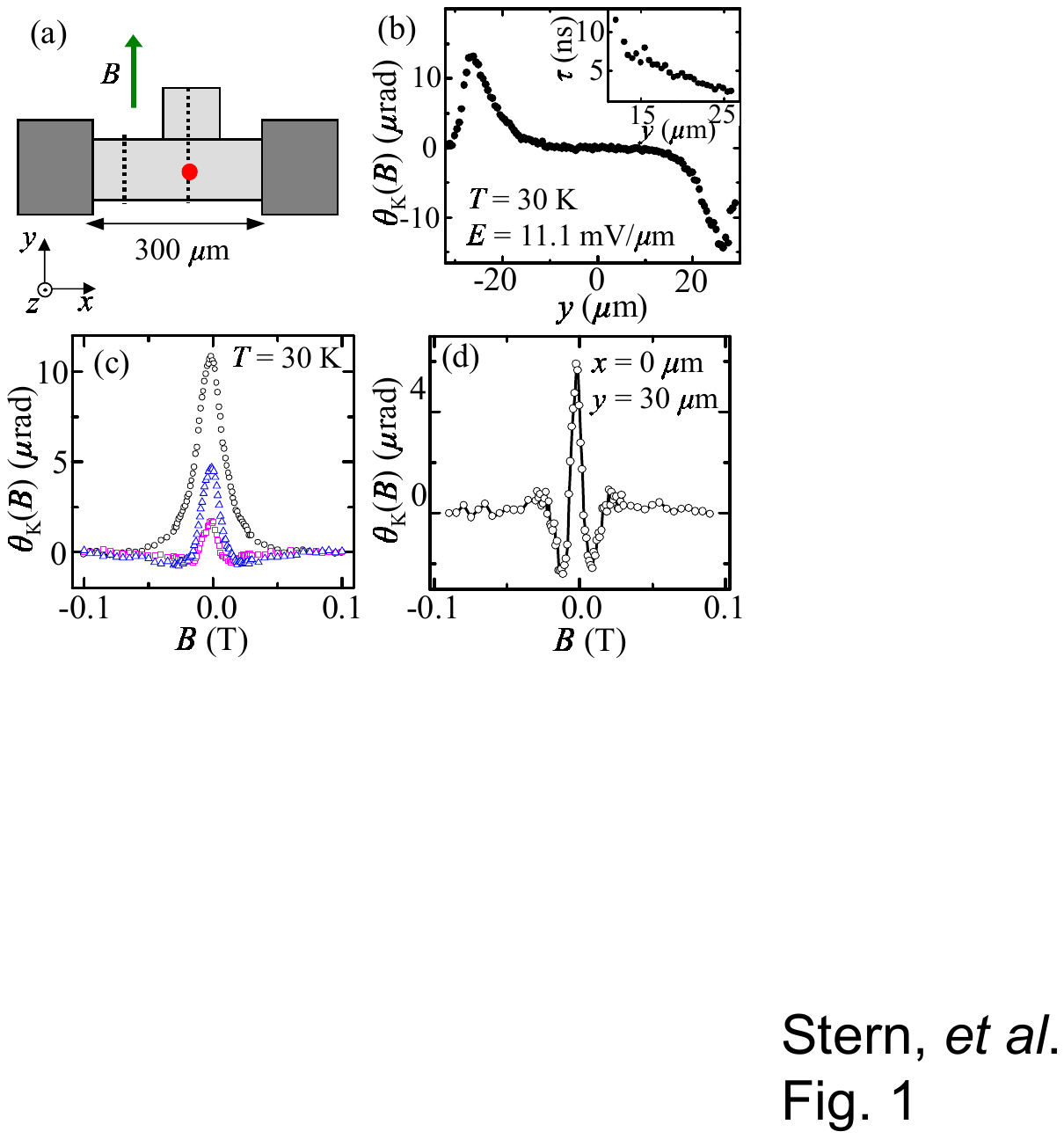}
  \end{center}
\end{figure}

\begin{figure}
  \begin{center}
    \includegraphics{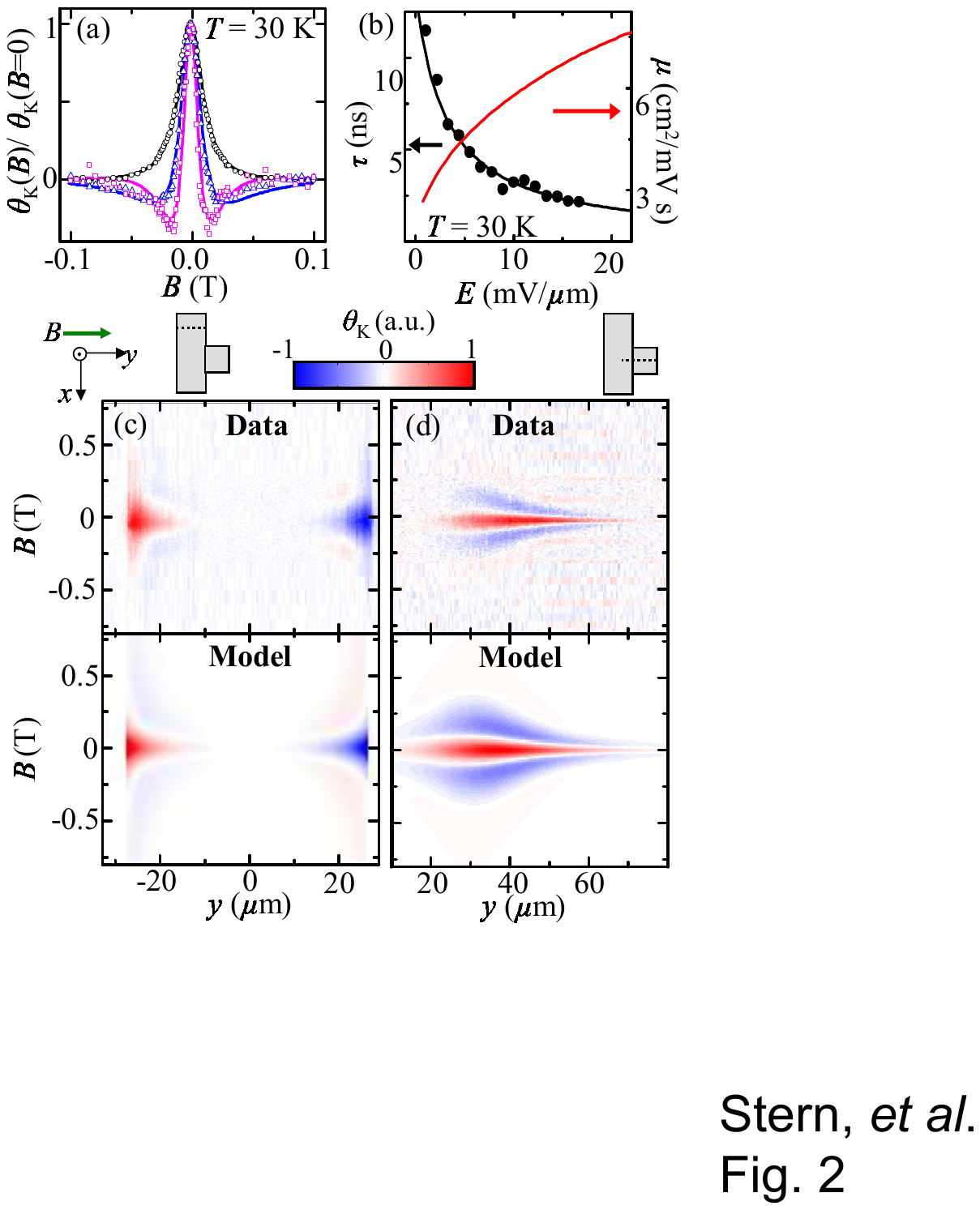}
  \end{center}
\end{figure}

\begin{figure}
  \begin{center}
    \includegraphics{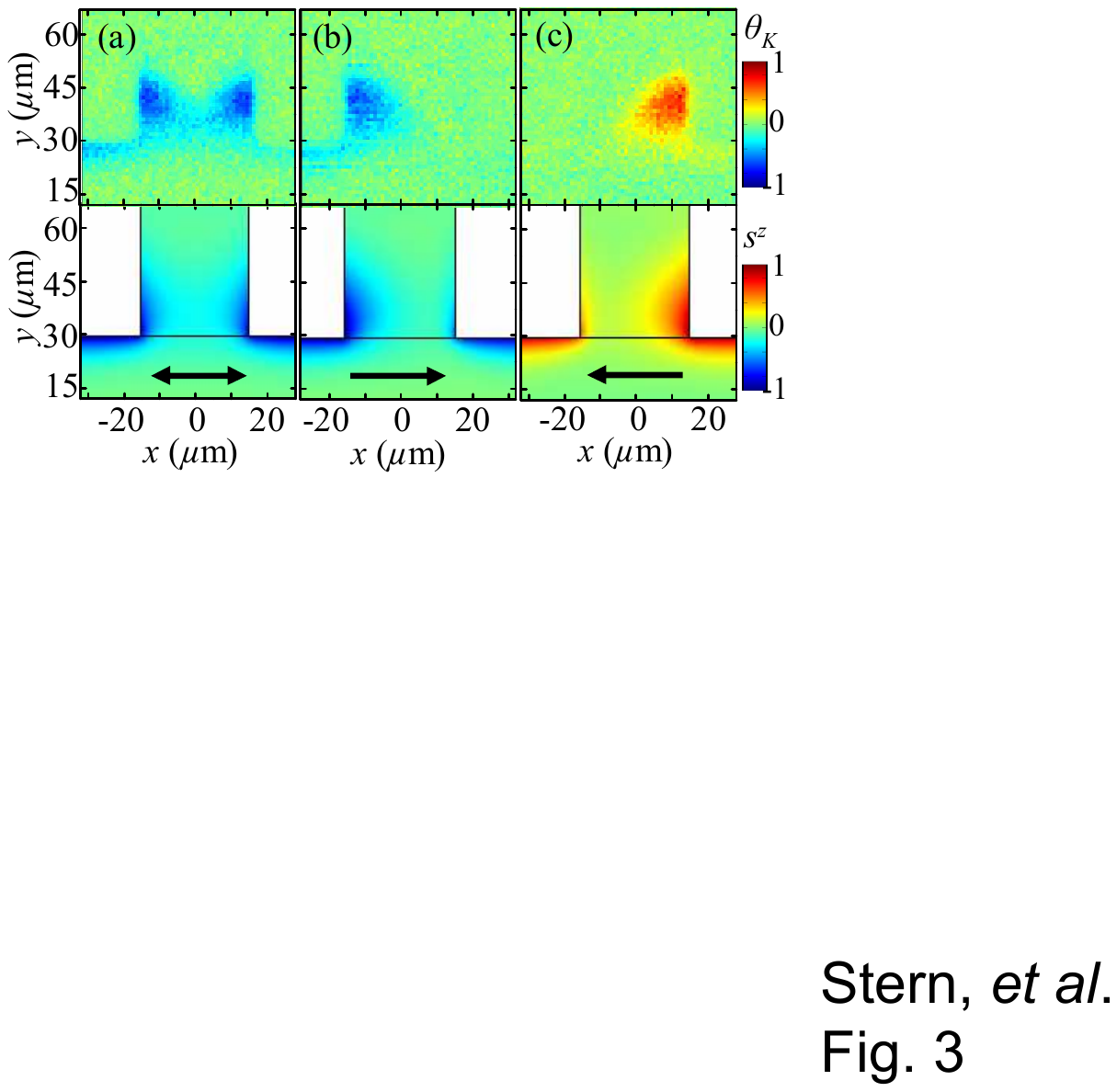}
  \end{center}
\end{figure}


\newpage

\begin{thebibliography}{16}

\bibitem{Kato:2004a} Y. Kato, R. C. Myers, A. C. Gossard, and D. D. Awschalom, Phys. Rev. Lett. \textbf{93}, 176601 (2004).
\bibitem{Kato:2004b} Y. Kato, R. C. Myers, A. C. Gossard, and D. D. Awschalom, Science \textbf{306}, 1910 (2004).
\bibitem{Sih:2006} V. Sih, W. H. Lau, R. C. Myers, V. R. Horowitz, A. C. Gossard, and D. D. Awschalom, Phys. Rev. Lett. \textbf{97}, 096605 (2006).
\bibitem{Shi:2006}  J. Shi, P. Zhang, D. Xiao, and Q. Niu, Phys. Rev. Lett. \textbf{96}, 076604 (2006).
\bibitem{Tse:2005} W-K. Tse, J. Fabian, I. \u{Z}uti\'{c}, and S. Das Sarma, Phys. Rev. B \textbf{72}, 241303(R) (2005).
\bibitem{Valenzuela:2006} S. O. Valenzuela and M. Tinkham, Nature \textbf{442}, 176 (2006).
\bibitem{Stern:2006} N. P. Stern, S. Ghosh, G. Xiang, M. Zhu, N. Samarth and D. D. Awschalom, Phys. Rev. Lett. \textbf{97}, 126603 (2006).
\bibitem{Stephens:2003} J. Stephens, R. K. Kawakami, J. Berezovsky, M. Hanson, D. P. Shepherd, A. C. Gossard, and D. D. Awschalom, Phys. Rev. B \textbf{68}, 041307(R) (2003).
\bibitem{Beck:2006} M. Beck, C. Metzner, S. Malzer, and G. H. D$\ddot{o}$hler, Europhys. Lett. \textbf{75}, 597 (2006).
\bibitem{Engel:2005} H-A. Engel, B. I. Halperin, and E. I. Rashba, Phys. Rev. Lett. \textbf{95}, 166605 (2005).
\bibitem{Finkler} I. Finkler, H-A Engel, E. I. Rashba, and B. I. Halperin, cond-mat/0703654.
\bibitem{Rashba:2006} E. I. Rashba, Physica E \textbf{34}, 31 (2006).
\bibitem{Johnson:1988} M. Johnson and R. H. Silsbee, Phys. Rev. B \textbf{37}, 5312 (1988).

\end{thebibliography}
\end{document}